\documentclass[journal,12pt,onecolumn,draftclsnofoot,]{IEEEtran}
\usepackage[ruled, vlined]{algorithm2e}
\usepackage{algpseudocode}
\usepackage{amsmath}
\usepackage{amssymb}
\usepackage{array}
\usepackage{arydshln}
\usepackage{blkarray, bigstrut}
\usepackage{bm}
\usepackage{booktabs}
\usepackage{braket}
\usepackage{cite}
\usepackage{cuted}
\usepackage{enumitem}
\usepackage{graphicx}
\usepackage[mathscr]{euscript}
\usepackage{mathtools}
\usepackage{multirow}
\usepackage{multicol}
\usepackage{framed} 
\usepackage[framed]{ntheorem}
\usepackage{nicefrac}
\usepackage{pgfplots}
\usepackage{qtree}
\usepackage{adjustbox}
\usepackage{rotating}
\pgfplotsset{compat = 1.13}
\usepackage{stfloats}
\usepackage{subfig}
\usepackage{url}
\usepackage{tikz}
\usepackage{tkz-berge}
\usepackage[framemethod=TikZ]{mdframed}
\usetikzlibrary{patterns}
\usetikzlibrary{spy}
\usetikzlibrary{shapes, backgrounds,calc, snakes}
\usetikzlibrary{decorations.pathreplacing}
\usetikzlibrary{matrix}
\usepackage{fancybox}

\tikzstyle{vertex} = [circle, draw, inner sep = 0pt, minimum size = 10pt]

\newframedtheorem{frm-prop}{Property}

\definecolor{bblue}{rgb}{0.12392, 0.0490, 0.9588}
\definecolor{sskyblue}{rgb}{0.1529, 0.5882, 0.9216}
\definecolor{ggreen}{rgb}{0.5020, 0.7961, 0.3451}
\definecolor{yyellow}{rgb}{0.9765, 0.9804, 0.0784}

\definecolor{color0}{HTML}{FF0147}
\definecolor{color1}{HTML}{F400DC}
\definecolor{color2}{HTML}{BA0DFF}
\definecolor{color3}{HTML}{5700E8}
\definecolor{color4}{HTML}{0B03FF}
\definecolor{color5}{HTML}{0957F4}
\definecolor{color6}{HTML}{03B3FF}
\definecolor{color7}{HTML}{08E8DA}
\definecolor{color8}{HTML}{07FF8E}
\definecolor{color9}{HTML}{51FF0A}

\definecolor{p1}{rgb}{1, 0.0667, 0}
\definecolor{p2}{rgb}{1, 0.24, 0}
\definecolor{p3}{rgb}{1, 0.349, 0}
\definecolor{p4}{rgb}{1, 0.490, 0}
\definecolor{p5}{rgb}{1, 0.631, 0}
\definecolor{p6}{rgb}{1, 0.792, 0}
\definecolor{p7}{rgb}{1, 0.933, 0}
\definecolor{p8}{rgb}{1, 1, 0}
\definecolor{p9}{rgb}{1, 1, 0.5}
\definecolor{p10}{rgb}{1, 1, 0.8}

\begin{document}

\title{TDOA-based Localization via Stochastic Gradient Descent Variants}

\author
{
	\IEEEauthorblockN{Luis F.~Abanto-Leon\IEEEauthorrefmark{2}, Arie Koppelaar\IEEEauthorrefmark{3}, Sonia Heemstra de Groot\IEEEauthorrefmark{4}} \\
	\IEEEauthorblockA{\IEEEauthorrefmark{2} \IEEEauthorrefmark{4} Eindhoven University of Technology}
	\IEEEauthorblockA{\IEEEauthorrefmark{3}NXP Semiconductors} \\ 
	\IEEEauthorrefmark{2}l.f.abanto@tue.nl,
	\IEEEauthorrefmark{3}arie.koppelaar@nxp.com,
	\IEEEauthorrefmark{4}sheemstradegroot@tue.nl,
}

\maketitle

\begin{abstract}
	Source localization is of pivotal importance in several areas such as wireless sensor networks and Internet of Things (IoT), where the location information can be used for a variety of purposes, e.g. surveillance, monitoring, tracking, etc. Time Difference of Arrival (TDOA) is one of the well-known localization approaches where the source broadcasts a signal and a number of receivers record the arriving time of the transmitted signal. By means of computing the time difference from various receivers, the source location can be estimated. On the other hand, in the recent few years novel optimization algorithms have appeared in the literature for $(i)$ processing big data and for $(ii)$ training deep neural networks. Most of these techniques are enhanced variants of the classical stochastic gradient descent (SGD) but with additional features that promote faster convergence. In this paper, we compare the performance of the classical SGD with the novel techniques mentioned above. In addition, we propose an optimization procedure called RMSProp+AF, which is based on RMSProp algorithm but with the advantage of incorporating adaptation of the decaying factor. We show through simulations that all of these techniques---which are commonly used in the machine learning domain---can also be successfully applied to signal processing problems and are capable of attaining improved convergence and stability. Finally, it is also shown through simulations that the proposed method can outperform other competing approaches as both its convergence and stability are superior.
\end{abstract}


\IEEEpeerreviewmaketitle

\section{Introduction}
Wireless sensor networks (WSN) and Internet of Things (IoT) are both constituted by assortments of interconnected devices with capabilities to $(i)$ communicate with each other and $(ii)$ interact with their surroundings. WSN are oftentimes regarded as a special case of IoT, where the devices are homogeneous and can only interface with specific types of devices. On the other hand, the premise in IoT is that devices could be based on heterogeneous technologies and yet they should be capable of collaborating and understanding each other in a smart manner. In both cases, the devices cooperate exchanging information with the aim of accomplishing a set of tasks. IoT and WSN  devices are found in a plethora of applications and across various domains. Among the several types of information that is utilized by the devices, location/position is of vital importance as it can be used for monitoring, tracking \cite{b1}, managing resources, etc. For instance, in the context of smart cities, the location can be used for identifying available parking spaces for vehicles. In the domain of Industry 4.0, such information can be exploited for improving the assignment of spectrum radio resources among the multiple sensors and actuators; thus leveraging communication reliability. Similarly, position data can be utilized by eNodeBs to enhance the distribution of resources in future vehicular networks \cite{b7}. In logistics and transportation, it is envisaged that real-time position data would be advantageous for transparent traceability of products, specially in complex manufacturing process \cite{b8}. In the area of healthcare, location data can be employed to track elderly people. Among the known techniques for estimating the location of a device, time difference of arrival (TDOA) has proved to be both robust and accurate. Furthermore, this technique is scalable and self-improvable, i.e. when more anchors (receivers) are available, these can be exploited to refine the position estimation and thereby mitigating measurement errors. 

On the other hand, several new optimization techniques have emerged in the domains of $(i)$ big data processing and $(i)$ deep neural networks (DNN) training. Most if not all of these techniques have their roots in stochastic gradient descent (SGD) but with additional integrated features that boost convergence and regularize stability. Among such techniques, stochastic gradient descent with momentum (SGD+M) \cite{b3}, root mean square propagation (RMSProp) \cite{b4} and adaptive moments (Adam) \cite{b5} are some of the most prominent approaches which can commonly outperform SGD in terms of convergence and accuracy. Form the standpoint of the authors, many of these techniques can be adapted to improve convergence in position estimation problem discussed herein.

\section{Motivation}
In order to provide real-time and accurate positioning, there is a necessity of counting with algorithms that are dependable and have fast convergence. This is particularly crucial in applications where position information is critical. For instance-, under the 4G/5G paradigms, cellular vehicle--to--everything (C-V2X) is a newly introduced technology that aims to support short-range vehicular communications. In C-V2X, vehicles broadcast beacon signals containing their location in order to provide awareness of its presence to neighboring vehicles and thus preventing accidents. However, in scenarios where global positioning system (GPS) signals are not available, location information will have to be derived from anchor points such as eNodeBs or roadside units (RSU). As a consequence, algorithms that can converge rapidly and provide accurate position information are of pivotal importance for this and other several applications.
 
In machine learning, training deep neural networks (DNN) with massive amounts of data is a commonly faced issue. Traditionally, SGD has been the mainstream method to accomplish such a purpose. However, the gradient in SGD may vary quickly after each iteration due to the wide variety of each input sample. Therefore, it is very likely that the steps converge slowly due to the chaotic gradient directions. Furthermore, SGD tends to have unstable update steps in each iteration and is prone to get trapped in poor local optima. In the recent few years, novel approaches based on SGD have emerged to address some of the main inconveniences of SGD. To alleviate the disadvantages of SGD, SDG+M was introduced. SDG+M incorporates knowledge from previous iterations and combines this information with the current gradient step to keep the updates stable. By means of this additional mechanism, the speed of learning is also boosted while chaotic jumps are prevented. RMSProp exponentially weights the past squared gradients by a forgetting factor in order to control the decaying learning rates. Therefore, RMSProp possesses outstanding stability and fast convergence. Adam constitutes an improvement of RMSProp. Adam incorporates momentum in the update rules, specifically for the first and second moments of the gradient. Additionally, Adam possess a bias correction procedure which compensates for the initial iterations where the estimations can be poorly predicted. However, due to the many integrated processes in Adam, convergence can sometimes be affected. 

Given the outstanding stability and accuracy of RMSProp, we propose a novel algorithm called RMSProp with adaptive decaying factor (RMSProp+AF) which adjusts an integrated exponential weight using knowledge from a short first-in first-out (FIFO) buffer containing previous gradients. By means of the proposed mechanism large steps are fostered at the beginning of the iterative process, which results in augmented convergence. Furthermore, as the iterative process progresses, the steps are reduced up to certain degree such that the learning rate does not diminish. We show that the proposed algorithm can outperform other state-of-the art techniques for the location estimation problem described in this paper.

The paper is structured as follows. In Section II, we define the TDOA model in a vectorized manner. Section III describes the proposed algorithm RMSProp+AF. Section IV is devoted to discussing simulation results. Finally, Section V summarizes the conclusions of our work.

\section{TDOA Model}
\begin{figure*}[!t]
	\begin{equation}\label{e1}
	\begin{array}{lll}
	\Delta \hat{\tau}_{ij} & = & \underset{\Delta \tau_{ij}}{\arg\max} \frac{\displaystyle \sum_{u}^{} \bar{z}_i(u) \bar{z}_j(u-\Delta \tau_{ij})}{\sqrt{\displaystyle \sum_{u} \bar{z}_i^2(u)} \sqrt{\displaystyle \sum_{u} \bar{z}_j^2(u)}} \\
	& = & \underset{\Delta \tau_{ij}}{\arg\max} 
	\frac{
		\displaystyle 
		\sum_{u} 
		\bigg( s(u-{\tau}_i) + \frac{h_i^{*}}{{\lvert h_i \lvert}^2} {\eta}_i(t) \bigg) 
		\bigg( s(u - \Delta \tau_{ij} -{\tau}_j) + \frac{h_j^{*}}{{\lvert h_j \lvert}^2} {\eta}_j(t) \bigg) 
	} 
	{
		\sqrt{\displaystyle \sum_{u} {\bigg( s(u - {\tau}_i) + \frac{h_i^{*}}{{\lvert h_i \lvert}^2} {\eta}_i(t) \bigg)}^2} 
		\sqrt{\displaystyle \sum_u {\bigg( s(u - {\tau}_j) + \frac{h_j^{*}}{{\lvert h_j \lvert}^2} {\eta}_j(t) \bigg)}^2}}  \\
	& = & ({\tau}_i-{\tau}_j) + {\pi}_{ij}
	\end{array}	
	\end{equation}
	\hrulefill
\end{figure*}

Consider a system consisting of a set of receivers (anchors) $\mathcal{R} = \{ r_1,r_2,\dots, r_N\}$ and a single transmitter whose position has to be estimated. In addition, the transmitter---at the unknown location $\bf p$---is actively broadcasting beacon signals $s(t)$ that are not necessarily known by the receivers. Furthermore, it is assumed that the receivers are located at known positions ${\bf \tilde{p}}_i = [ \tilde{x}_i, \tilde{y}_i ]^T$, $i = 1, 2, \cdots, N$. Let $z_i(t) = h_i \cdot s(t-{\tau}_i) + \eta_i(t)$ denote the received signal at receiver $r_i \in \mathcal{R}$. The parameter ${\tau}_i$ represents the time of arrival at the receiver $r_i$. The channel gain at receiver $r_i$ is denoted by $h_i$ whereas ${\eta}_i $ represents Gaussian noise. When the transmitted signal $s(t)$ is unknown by the receivers, a viable idea is to remove the incognito signal $s(t)$ by means of correlation analysis \cite{b1}. Thus, the TDOA measurements $\Delta \tau_{ij}$ are indirectly estimated by computing the normalized cross-correlation (NCC) between every pair of signals $\bar{z}_i(t)$ and $\bar{z}_j(t)$. To wit, the objective is to find an estimate $\Delta \hat{\tau}_{ij}$ of the true value $\Delta \tau_{ij}$ via NCC, as shown in (\ref{e1}).

In (\ref{e1}), ${\pi}_{ij}$ is a systematic error measurement with Gaussian distribution. The signals $\bar{z}_i(t) = \frac{h_i^{*}}{{\lvert h_i \lvert}^2} z_i(t) = s(t-{\tau}_i) + \frac{h_i^{*}}{{\lvert h_i \lvert}^2} {\eta}_i(t)$ and $\bar{z}_j(t) = \frac{h_j^{*}}{{\lvert h_j \lvert}^2} z_j(t) = s(t-{\tau}_j) + \frac{h_j^{*}}{{\lvert h_j \lvert}^2} {\eta}_j(t)$ are equalized versions of the received $z_i(t)$ and $z_j(t)$. Although the TDOA estimate $\Delta \hat{\tau}_{ij}$ is computed from the equalized received signals, the NCC was considered instead of the conventional cross-correlation due its robustness to gain difference. By exploiting the received signals in a pair-wise manner, both the unknown transmit time $t$ and the unknown signal $s(t)$ can be removed. Because the underlying location estimation problem requires using distances, all TDOAs $\Delta \hat{\tau}_{ij}$ will be converted from time to range differences as shown in (\ref{e2}). 
\begin{equation} \label{e2}
	\begin{array}{lll}
		\Delta \hat{d}_{ij}  & = & c \cdot \Delta \hat{\tau}_{ij} \\
		                & = & c \cdot ({\tau}_i-{\tau}_j) + c \cdot {\pi}_{ij} \\
						& = & d_i - d_j + {\epsilon}_{ij} \\
						& = & {\| {\bf p} - {\bf \tilde{p}}_i \|}_2 - {\| {\bf p} - {\bf \tilde{p}}_j \|}_2 + {\epsilon}_{ij} \\
						& = & g({\bf p}, {\bf \tilde{p}}_i, {\bf \tilde{p}}_j) + {\epsilon}_{ij}
	\end{array}
\end{equation}
with ($1 \leq i < j \leq N$) and ${\epsilon}_{ij} \sim \mathcal{N} (0, {\sigma}^2_i + {\sigma}^2_j)$. Note that each $\Delta \hat{d}_{i,j}$ corresponds to a set of potential positions ${\bf p} = [x~y]^T$ along a hyperbola where the transmitter could be located. It is therefore necessary at least three anchors in order to obtain an intersection of two hyperbolas. In addition, note that $\Delta d_{ij}$ and $\Delta d_{ji}$ will define the same hyperbola. Thus, for notation simplification
\begin{equation} \label{e3}
	\Delta \hat{d}_{m} = g({\bf p}, \tilde{{\bf p}}_i, \tilde{{\bf p}}_j) + {\epsilon}_{m}
\end{equation}
where $1 \leq m \leq M$, $M = \frac{N!}{2!(N-2)!}$ and ${\epsilon}_m \sim \mathcal{N} (0, {\sigma}^2_m)$. In this paper, the terms \textit{TDOA} and \textit{range difference} will be employed interchangeably \cite{b6}. Also, note that \textit{range difference} between two receivers is with respect to the transmitter. 
\begin{figure*}[!t]
	\begin{equation} \label{e4}
	p({\bf \Delta \hat{d}} \mid {\bf p}) = \frac{1}{\sqrt{\det (2 \pi {\bf C})}} \exp \Big(-\frac{1}{2} ({\bf \Delta \hat{d} - g})^T{\bf C}^{-1}({\bf \Delta \hat{d} - g}) \Big)
	\end{equation}
	\hrulefill
\end{figure*}

Given the observed measurements ${\bf \Delta \hat{d}}= [{\Delta} \hat{d}_{1,2}, {\Delta} \hat{d}_{1,3}, \cdots, {\Delta} \hat{d}_{N-1,N}]^T$, the objective is to estimate---with the least uncertainty---the true position $\bf p$ of the transmitter. This problem can be formulated as maximizing the likelihood function \cite{b2} shown in (\ref{e3}), where $\bf C$ is the covaraince matrix of the measurement errors. The vector $\bf g$ is given by (\ref{e4})
\begin{equation} \label{e5}
	{\bf g} = 
	\left[ 
		\begin{array}{c} 
			{\| {\bf p} - \tilde{{\bf p}}_1 \|}_2 - {\| {\bf p} - \tilde{{\bf p}}_2 \|}_2 \\
			{\| {\bf p} - \tilde{{\bf p}}_1 \|}_2 - {\| {\bf p} - \tilde{{\bf p}}_3 \|}_2\\
			\vdots \\
			{\| {\bf p} - \tilde{{\bf p}}_{N-1} \|}_2 - {\| {\bf p} - \tilde{{\bf p}}_N \|}_2
		\end{array} 
	\right]. 
\end{equation}

To wit, maximizing (\ref{e3}) is equivalent to minimizing (\ref{e6})
\begin{equation} \label{e6}
	{\bf \hat{p}} = \arg \min_{\bf p} \underbrace{({\bf \Delta \hat{d} - g})^T{\bf C}^{-1}({\bf \Delta \hat{d} - g})}_{J:~\text{cost function}}.
\end{equation}

In order to determine $\bf p$, we opt for using a gradient approach where an initial position estimate is iteratively updated until a near-optimal solution is obtained. To this purpose, the gradient of $J$ with respect to $\bf p$, i.e. ${\nabla}_{\bf p} J$ is computed and the position is updated iteratively. Thus, in the case of SGD, the estimated location at iteration $k$ is expressed by  
\begin{equation} \label{e7}
	{\bf \hat{p}}^{(k+1)} = {\bf \hat{p}}^{(k)} - \mu {\nabla}_{\bf p}^{(k)} J, 
\end{equation}
where $\mu$ is the learning rate and ${\nabla}_{\bf p}^{(k)} J$ is given by
\begin{equation} \label{e8}
	{\nabla}_{\bf p}^{(k)} J = -2 {\boldsymbol \epsilon}^{(k)}
	\left[ 
		\begin{array}{c} 
			\frac{{\bf p}^{(k)} - \tilde{{\bf p}}_1}{{\| {\bf p}^{(k)} - \tilde{{\bf p}}_1 \|}_2} - \frac{{\bf p}^{(k)} - \tilde{{\bf p}}_2}{{\| {\bf p}^{(k)} - \tilde{{\bf p}}_2 \|}_2} \\
			\frac{{\bf p}^{(k)} - \tilde{{\bf p}}_1}{{\| {\bf p}^{(k)} - \tilde{{\bf p}}_1 \|}_2} - \frac{{\bf p}^{(k)} - \tilde{{\bf p}}_3}{{\| {\bf p}^{(k)} - \tilde{{\bf p}}_3 \|}_2} \\
			\vdots \\
			\frac{{\bf p}^{(k)} - \tilde{{\bf p}}_{N-1}}{{\| {\bf p}^{(k)} - \tilde{{\bf p}}_{N-1} \|}_2} - \frac{{\bf p}^{(k)} - \tilde{{\bf p}}_N}{{\| {\bf p}^{(k)} - \tilde{{\bf p}}_N \|}_2} \\
		\end{array} 
	\right] 
\end{equation}
with ${\boldsymbol \epsilon}^{(k)} = {\bf \Delta d} - {\bf g}^{(k)} $ denoting the estimation error vector at the $k$-th iteration.

\section{Proposed RMSProp+AF}
The proposed optimization approach based on RMSProp is described in great detail in Algorithm \ref{a1}. From \textit{Step 1} to \textit{Step 4}, useful variables are initialized. For instance, the buffers $b_x$ and $b_y$ of size $L$ are both emptied. In addition, the decaying factors thresholds contained in ${\boldsymbol \rho}^{(0)}$ are predefined. Also, the canonical vectors for two dimensions, ${\bf u}_x$ and ${\bf u}_y$ are specified. In \textit{Step 5a}, an index $k'$ is computed to emulate the FIFO buffers $b_x$ and $b_y$. In \textit{Step 5b}, the square of the gradient is calculated and each of the resultant components is stored in the buffer it corresponds to. Realize that the buffers will preserve the $L$ most recent squared gradients, which will be utilized in the upcoming processes. In \textit{Step 5c}, two auxiliary vectors ${\bf v}_{max}$ and ${\bf v}_{min}$ are computed. Each of these vectors will respectively contain the maximum and minimum values of $b_x$ and $b_y$, i.e. the maximum and minimum squared gradients per dimension among the most recent $L$. These will be used for updating the decaying weighting factor at each iteration. As can be observed there is a decaying factor per dimensions and each is independent. Finally, \textit{Step 5e} and \textit{Step 5f} are the update equations originally stated in RMSProp. To wit, \textit{Step 5e} provides a smoothed version of the squared gradient whereas \textit{Step 5f} applies a power normalization. \textit{Step 6} gives the final estimation of the position.
\begin{center}
	\begin{table}[!b]
		\centering
		\scriptsize
		\caption {Simulation parameters}
		\label{t1}
		\begin{tabular}{clcc}
			\toprule
			\multicolumn{1}{c}{\textbf{Algorithm}} & \multicolumn{1}{c}{\textbf{Parameter}} & \multicolumn{1}{c}{\textbf{Symbol}} & \multicolumn{1}{c}{\textbf{Value}} \\
			\midrule
			\multirow{1}{*}{SGD} & Learning rate  & $\mu$    & 0.01    \\										 
			\midrule
			\multirow{2}{*}{SGD+M} & Learning rate  & $\mu$    & 0.01 \\
								   & Momentum term & $\alpha$  & 0.9 \\
			\midrule
			\multirow{3}{*}{RMSProp} & Learning rate  & $\mu$    & 0.01 \\
								     & Decaying factor & $\rho$  & 0.999 \\		
								     & Smoothing term & $\delta$  & 0.000001 \\				   
			\midrule
			\multirow{4}{*}{Adam} & Learning rate  & $\mu$    & 0.01 \\
								  & Decaying factor 1 & ${\rho}_1$  & 0.9 \\
								  & Decaying factor 2 & ${\rho}_2$  & 0.999 \\		
								  & Smoothing term & $\delta$  & 0.000001 \\	
			\midrule
			\multirow{3}{*}{RMSProp+AF} & Learning rate  & $\mu$    & 0.01 \\
									   & Decaying factor threshold & $\rho$  & 0.99 \\		
									   & Smoothing term & $\delta$  & 0.000001 \\	
			
			\bottomrule
		\end{tabular}
	\end{table}
	\vspace{-0.75cm}
\end{center}
\begin{algorithm*}
	\DontPrintSemicolon
	
	\KwIn{ The gradient $\nabla_{\bf p}^{(k)} J$ of cost function $J$} 
	\KwOut{The estimated position ${\bf \hat{p}}$}
	
	\Begin
	{
		\begin{tabular}{m{0.7cm} m{12cm}}
			\underline{\textit{Step 1:}} & Initialize the FIFO buffers $b_x$ and $b_y$ of size $L$. \\
			\underline{\textit{Step 2:}} & Initialize  ${\boldsymbol \rho}^{(0)} = [0.99~0.99]^T$. \\
			\underline{\textit{Step 3:}} & Initialize  ${\bf r}^{(0)} = [0~0]^T$. \\
			\underline{\textit{Step 4:}} & Define the vectors ${\bf u}_x = [1~0]^T$ and ${\bf u}_y = [0~1]^T$. \\
			
										 &                                        \\
		\end{tabular}

		\For{$k = 1: K$}
		{	
			\begin{tabular}{m{1.0cm} m{12cm}}
				\underline{\textit{Step 5a:}} & Compute the circular buffer index $k'$ \\
											  & $k' = k-L\left\lfloor{\displaystyle \frac{k-1}{L}}\right\rfloor$ \\
				\underline{\textit{Step 5b:}} & Store the square of gradient $\nabla_{\bf p}^{(k)} J$ at index $k'$ on each of the buffers \\ 						 
											  & $b_x(k') = {\bf u}_x^T \Big(\nabla_{\bf p}^{(k)} J \odot \nabla_{\bf p}^{(k)} J \Big) $ \\
											  & $b_y(k') = {\bf u}_y^T \Big(\nabla_{\bf p}^{(k)} J \odot \nabla_{\bf p}^{(k)} J \Big) $ \\
				\underline{\textit{Step 5c:}} & Define the vectors ${\bf v}_{max}$ and ${\bf v}_{min}$ \\
		 							          & ${\bf v}_{max} = \begin{bmatrix} \max \{b_x\} \\ \max \{b_y\} \end{bmatrix},  {\bf v}_{min} \begin{bmatrix} \min \{b_x\} \\ \min \{b_y\} \end{bmatrix}$ \\
		 		\underline{\textit{Step 5d:}} & Compute the adaptive decaying factor ${\boldsymbol \rho}^{(k)}$ \\
		 							          & $ {\boldsymbol \gamma}^{(k)} = ({\bf v}_{max} - {\bf v}_{min}) \oslash ({\bf v}_{max} + {\bf v}_{min} + {\bf 1}_{2 \times 1}) $ \\
		 							          & ${\boldsymbol \rho}^{(k)} = \begin{bmatrix} \max \big \{ {\bf u}_x^T {\boldsymbol \rho}^{(0)}, {\bf u}_x^T {\boldsymbol \gamma}^{(k)} \big \} \\ \max \big \{ {\bf u}_y^T {\boldsymbol \rho}^{(0)}, {\bf u}_y^T {\boldsymbol \gamma}^{(k)} \big \} \end{bmatrix} $\\   
    			\underline{\textit{Step 5e:}}  & Accumulate the squared gradient \\
		    							      & ${\bf r}^{(k)} = {\boldsymbol \rho}^{(k)} \odot {\bf r}^{(k-1)}  + ({\bf 1}_{2 \times 1} - {\boldsymbol \rho}^{(k)}) \odot \nabla_p^{(k)} J \odot \nabla_p^{(k)} J $\\
				\underline{\textit{Step 5f:}}  & Update the position \\
										      & ${\bf p}^{(k+1)} = {\bf p}^{(k)} - \frac{\mu}{\delta+\sqrt{{\bf r}^{(k)}}} \odot {\nabla}_{\bf p} J$
			\end{tabular}
		}
	
		\begin{tabular}{m{0.7cm} m{12cm}}
			\underline{\textit{Step 6:}} & Output ${\bf \hat{p}} ={\bf p}^{(K+1)} $ \\
			&                                        \\
		\end{tabular}
	}
	\caption{ Proposed RMSProp with Adaptive Decaying Factor (RMSProp+AF) }
	\label{a1}
\end{algorithm*}

\section{Simulations}
In this section, we evaluate two types of scenario. In the first case, the device---whose location has to be estimated---is inside the convex hull defined by the $N$ receivers. In the second scenario, the device is located outside the convex hull. As it is known, the second scenario is more challenging and therefore the estimation error is expected to be higher. For both scenarios, the parameters of the algorithms have been specified in Table \ref{t1}.

\underline{\textit{Scenario 1}}: \emph{Consider that $N = 4 $ receivers are located at positions ${\bf \tilde{p}}_1 = [0~0]^T$, ${\bf \tilde{p}}_2 = [10~60]^T$, ${\bf \tilde{p}}_3 = [70~70]^T$ and ${\bf \tilde{p}}_4 = [60~10]^T$. In addition, the unknown position of the transmitter is ${\bf {p}} = [40~80]^T$.}
 
The described scenario is depicted in Fig. \ref{f1}. We have considered a covariance matrix with diagonal and off-diagonal elements equal to 0.4 and 0.1, respectively. For this reason, the true position may not be found correctly and all the approaches will converge to a slightly shifted solution. In Fig. \ref{f1} we observe that the proposed approach RMSProp+AF exhibits large steps at the beginning while progressively reduces the step-size as the process advances. This is a consequence of performing adaptation of the decaying factor. We also observe that SGD+M exhibits consistent step-sizes and has higher convergence rate than other competing algorithms. Based on Fig. \ref{f2}, we can realize that RMSProp+AF can---within a few iterations---attain a near-optimal solution compared to other approaches. For instance, the estimation error after 50 iterations is approximately 3 meters. SGD+M requires as much as twice iterations to reach stability and comparable error as RMSProp+AF. On the other hand, RMSProp requires more than 300 iterations to reach a similar estimation error. Both SGD and Adam exhibit slow convergence and are outperformed by the previously mentioned algorithms. Counterintuitively, the highly-elaborated Adam is outperformed by the simplistic SGD. This might be due to ineffective biased correction and the extra parameters that do not work harmoniously. 
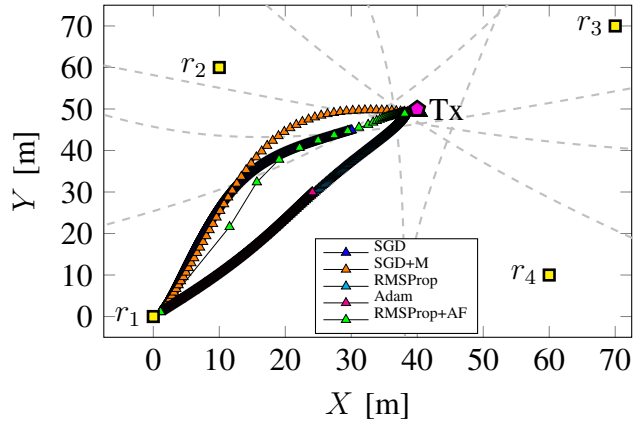
\begin{figure}[!t] 
	\centering
	\begin{tikzpicture}
	\begin{axis}
	[
	xlabel={$X$ [m]},
	ylabel = {$Y$ [m]},
	xmin=-0.75,
	xmax=7.25, 
	ymin=-0.50, 
	ymax=7.50,
	xtick = {0, 1, 2, 3, 4, 5, 6, 7},
	xticklabels = {0, 10, 20, 30, 40, 50, 60, 70},
	ytick = {0, 1, 2, 3, 4, 5, 6, 7},
	yticklabels = {0, 10, 20, 30, 40, 50, 60, 70},
	height = 6cm,
	width = 8.6cm,
	legend columns=1,
	legend pos = north east,
	legend style={at={(0.4,0.01)},anchor=south west, font=\fontsize{6}{5}\selectfont, text width=1.2cm,text height=0.02cm,text depth=.ex, fill = white, align = left},
	]
	\addplot[color=black, mark = triangle*, mark options = {fill = color4, solid}, line width = 0.1pt] table {SGD.txt}; \addlegendentry{SGD}
	
	\addplot[color=black, mark = triangle*, mark options = {fill = orange, solid}, line width = 0.1pt] table {SGDM.txt}; \addlegendentry{SGD+M}
	
	\addplot[color=black, mark = triangle*, mark options = {fill = cyan, solid}, line width = 0.1pt] table {RMSProp.txt}; \addlegendentry{RMSProp}
	
	\addplot[color=black, mark = triangle*, mark options = {fill = magenta, solid}, line width = 0.1pt] table {Adam.txt}; \addlegendentry{Adam}
	
	\addplot[color=black, mark = triangle*, mark options = {fill = green, solid}, line width = 0.1pt] table {RMSPropAF.txt}; \addlegendentry{RMSProp+AF}
	
	\addplot [color=lightgray, mark options = {scale = 0.8, fill = blue}, line width = 0.8pt, dashed] table {Hyperbola1.txt};
	
	\addplot [color=lightgray, mark options = {scale = 0.8, fill = blue}, line width = 0.8pt, dashed] table {Hyperbola2.txt};
	
	\addplot [color=lightgray, mark options = {scale = 0.8, fill = blue}, line width = 0.8pt, dashed] table {Hyperbola3.txt};
	
	\addplot [color=lightgray, mark options = {scale = 0.8, fill = blue}, line width = 0.8pt, dashed] table {Hyperbola4.txt};
	
	\addplot [color=lightgray, mark options = {scale = 0.8, fill = blue}, line width = 0.8pt, dashed] table {Hyperbola5.txt};
	
	\addplot [color=lightgray, mark options = {scale = 0.8, fill = blue}, line width = 0.8pt, dashed] table {Hyperbola6.txt};
	
	\addplot[color=black, mark = square*, mark options = {fill = yellow, solid}, line width = 1pt] coordinates {(0,0)}  node[left,pos=1] {$r_1$}; \label{p4} 
	
	\addplot[color=black, mark = square*, mark options = {fill = yellow, solid}, line width = 1pt] coordinates {(1,6)}  node[left,pos=1] {$r_2$};
	
	\addplot[color=black, mark = square*, mark options = {fill = yellow, solid}, line width = 1pt] coordinates {(7,7)}  node[left,pos=1] {$r_3$}; 
	
	\addplot[color=black, mark = square*, mark options = {fill = yellow, solid}, line width = 1pt] coordinates {(6,1)}  node[left,pos=1] {$r_4$}; 
	
	\addplot[align=center, color = black, mark = pentagon*, mark options = {scale = 1.5, fill = color1, solid}, line width = 1pt] coordinates {(4,5)} node[right,pos=1] {Tx}; 
	
	\end{axis}
	
	\end{tikzpicture}
	\caption{Iterative process for position estimation}
	\label{f1}
\end{figure}
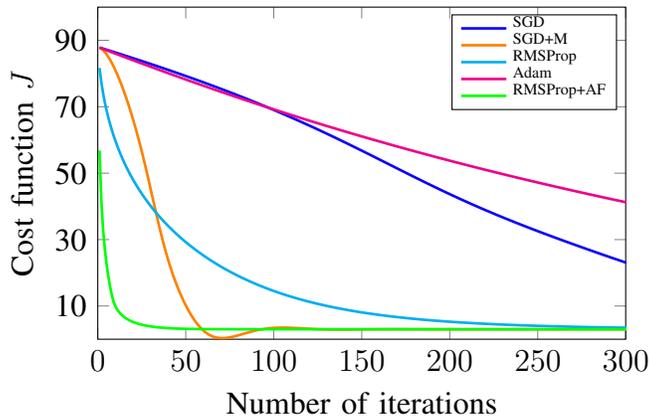
\begin{figure}[!t]
	\centering
	\begin{tikzpicture}
	\begin{axis}
	[
	xlabel={Number of iterations},
	ylabel = {Cost function $J$},
	xmin=0,
	xmax=300, 
	ymin=0, 
	ymax=10,
	ytick = {1, 3, 5, 7, 9},
	yticklabels = {10, 30, 50, 70, 90},
	width = 8.6cm,
	height = 6cm,
	legend columns=1,
	legend pos = north east,
	legend style={at={(0.67,0.69)},anchor=south west, font=\fontsize{6}{5}\selectfont, text width=1.2cm,text height=0.02cm,text depth=.ex, fill = white, align = left},
	]
	\addplot[color=color4, mark = none, line width = 1pt] table {errorSGD.txt}; \addlegendentry{SGD}
	
	\addplot[color=orange, mark = none, line width = 1pt] table {errorSGDM.txt}; \addlegendentry{SGD+M}
	
	\addplot[color=cyan, mark = none, line width = 1pt] table {errorRMSProp.txt}; \addlegendentry{RMSProp}
	
	\addplot[color=magenta, mark = none, line width = 1pt] table {errorAdam.txt}; \addlegendentry{Adam}
	
	\addplot[color=green, mark = none, line width = 1pt] table {errorRMSPropAF.txt}; \addlegendentry{RMSProp+AF}
		
	\end{axis}
	\end{tikzpicture}
	\caption{Scenario 1 - Convergence of algorithms}
	\label{f2}
\end{figure}

\underline{\textit{Scenario 2}}: \emph{Consider that $N = 4 $ receivers are located at positions ${\bf \tilde{p}}_1 = [0~0]^T$, ${\bf \tilde{p}}_2 = [10~60]^T$, ${\bf \tilde{p}}_3 = [70~70]^T$ and ${\bf \tilde{p}}_4 = [60~10]^T$. In addition, the unknown position of the transmitter is ${\bf {p}} = [75~65]^T$.}
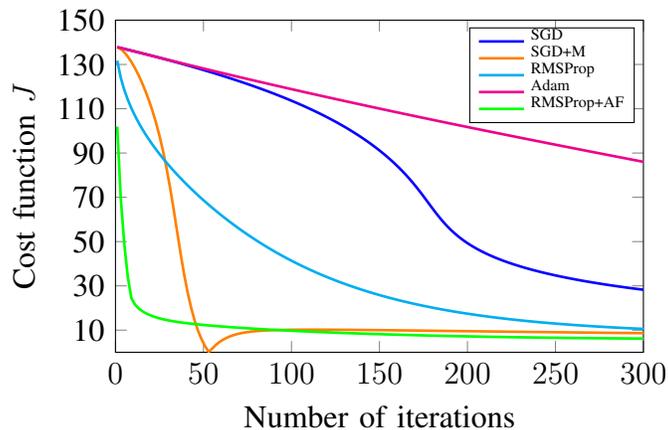
\begin{figure}[!t]
	\centering
	\begin{tikzpicture}
	\begin{axis}
	[
	xlabel={Number of iterations},
	ylabel = {Cost function $J$},
	xmin=0,
	xmax=300, 
	ymin=0, 
	ymax=15,
	ytick = {1, 3, 5, 7, 9, 11, 13, 15},
	yticklabels = {10, 30, 50, 70, 90, 110, 130, 150},
	width = 8.6cm,
	height = 6cm,
	legend columns=1,
	legend pos = north east,
	legend style={at={(0.67,0.69)},anchor=south west, font=\fontsize{6}{5}\selectfont, text width=1.2cm,text height=0.02cm,text depth=.ex, fill = white, align = left},
	]
	\addplot[color=color4, mark = none, line width = 1pt] table {errorSGD2.txt}; \addlegendentry{SGD}
	
	\addplot[color=orange, mark = none, line width = 1pt] table {errorSGDM2.txt}; \addlegendentry{SGD+M}
	
	\addplot[color=cyan, mark = none, line width = 1pt] table {errorRMSProp2.txt}; \addlegendentry{RMSProp}
	
	\addplot[color=magenta, mark = none, line width = 1pt] table {errorAdam2.txt}; \addlegendentry{Adam}
	
	\addplot[color=green, mark = none, line width = 1pt] table {errorRMSPropAF2.txt}; \addlegendentry{RMSProp+AF}
	
	\end{axis}
	\end{tikzpicture}
	\caption{Scenario 2 - Convergence of algorithms}
	\label{f3}
\end{figure}

We observe in Fig. \ref{f3} that RMSProp+AF is the best-performing among all the assessed algorithms. RMSProp is capable of attaining similar error performance as RMSProp+AF within 150 iterations. However, RMSProp+AF does not exhibit oscillations and it is therefore more stable. Because the transmitter in not located inside the convex hull, the estimation error is higher compared to the first case; approximately 6 meters when using RMSProp+AF. The error floor using RMSProp is 9 meters. The remaining three algorithms exhibit very slow convergence although the hyper-parameters have been tuned the fairest possible. 

\section{Conclusion}
In this work we have presented a comparison of different optimization techniques---commonly used in the machine learning realm---to solve TDOA-based localization. We can conclude that most of the approaches can be successfully applied and can outperform classical methods such as stochastic gradient descent. In addition, we presented an improved version named RMSProp+AF, which is capable of providing enhanced convergence in comparison to state--of--the--art approaches. We showed through simulations, that the proposed scheme outperforms other competing approaches in two types of scenarios, i.e. $(i)$ when the transmitter is inside and $(ii)$ when it is outside the convex hull formed by the receivers.

\end{document}